\documentclass[conference]{IEEEtran}
\ifCLASSOPTIONcompsoc
\else
\fi
\ifCLASSINFOpdf
\else
\fi
\usepackage{amsmath}
\usepackage[english]{babel}
\usepackage[parfill]{parskip}
\usepackage{changes}
\usepackage[style=ieee, backend=bibtex, minnames=1]{biblatex} 

\DeclareBibliographyCategory{cited}
\AtEveryCitekey{\addtocategory{cited}{\thefield{entrykey}}}
\usepackage{pdflscape}
\usepackage[toc,acronym]{glossaries}
\usepackage{todonotes}

\usepackage{enumitem}
\usepackage{booktabs}

\usepackage{tikz,tkz-tab}
\usetikzlibrary{babel}
\usepackage{circuitikz}
\usepackage{tikzpeople}
\usepackage{modules/tikzNetwork}
\usetikzlibrary{positioning, automata, graphs, trees, fit, arrows.meta, shapes}
\usetikzlibrary{backgrounds,patterns,matrix,calc,shadows,plotmarks, circuits.logic.US}
\usetikzlibrary{decorations}
\usepackage{pgffor}
\usepackage{multirow}
\usepackage{pgfgantt}
\usepackage{modules/moeptikz}
\input{modules/circuitDesignLibrary}
\usepackage{cancel}
\usepackage{amsmath}
\usepackage{stmaryrd}
\usepackage{textcomp}
\usepackage{pstricks}
\usepackage{pdftricks, pstricks-add}
\usepackage{pgfplots}
\usepackage{caption}
\usepackage{subfig}
\usepackage{pdfpages}
\usepackage[nounderscore]{syntax}
\input{modules/grammarSpec.tex}
\usepackage{varwidth}

\usepackage{tabularx}
\usepackage{csvsimple}
\usepackage{float}
\usepackage{amssymb}
\usepackage[linesnumbered,lined, french, frenchkw, figure, noend]{algorithm2e}

\usepackage{listings}

\lstdefinelanguage{p4}
{ morekeywords={*,extern_type, extern, table, apply, typedef, header, attribute, type, method, extern, action, control, void, parser,state, start, transition, extract, select, reject, default, accept, out, in, inout, return, bit, package},
	sensitive=true,
	morecomment=[l]{//}, 
	morecomment=[s]{/*}{*/}, 
	morestring=[b]" 
}
\lstdefinelanguage{p414}
{ morekeywords={*,extern_type, attribute, type, method, extern, action, control, void, parser,state, start, transition, extract, select, default, accept, out, in, inout, return},
	sensitive=true,
	morecomment=[l]{//}, 
	morecomment=[s]{/*}{*/}, 
	morestring=[b]" 
}

\tikzstyle{vecArrow} = [thick, decoration={markings,mark=at position
	1 with {\arrow[semithick]{open triangle 60}}},
double distance=1.4pt, shorten >= 5.5pt,
preaction = {decorate},
postaction = {draw,line width=1.4pt, white,shorten >= 4.5pt}]

\newcommand{\inputFig}[2]{\IfFileExists{#1}{\input{#1}}{\missingfigure[figheight=3cm]{#2}}} 

\makeatletter

\pgfdeclareshape{dff}{
  \savedanchor\northeast{%
    \pgfmathsetlength\pgf@x{\pgfshapeminwidth}%
    \pgfmathsetlength\pgf@y{\pgfshapeminheight}%
    \pgf@x=0.5\pgf@x
    \pgf@y=0.5\pgf@y
  }
  \savedanchor\southwest{%
    \pgfmathsetlength\pgf@x{\pgfshapeminwidth}%
    \pgfmathsetlength\pgf@y{\pgfshapeminheight}%
    \pgf@x=-0.5\pgf@x
    \pgf@y=-0.5\pgf@y
  }
  \inheritanchorborder[from=rectangle]

  \anchor{center}{\pgfpointorigin}
  \anchor{north}{\northeast \pgf@x=0pt}
  \anchor{east}{\northeast \pgf@y=0pt}
  \anchor{south}{\southwest \pgf@x=0pt}
  \anchor{west}{\southwest \pgf@y=0pt}
  \anchor{north east}{\northeast}
  \anchor{north west}{\northeast \pgf@x=-\pgf@x}
  \anchor{south west}{\southwest}
  \anchor{south east}{\southwest \pgf@x=-\pgf@x}
  \anchor{text}{
    \pgfpointorigin
    \advance\pgf@x by -.5\wd\pgfnodeparttextbox%
    \advance\pgf@y by -.5\ht\pgfnodeparttextbox%
    \advance\pgf@y by +.5\dp\pgfnodeparttextbox%
  }

  \anchor{D}{
    \pgf@process{\northeast}%
    \pgf@x=-1\pgf@x%
    \pgf@y=.5\pgf@y%
  }
  \anchor{CLK}{
    \pgf@process{\northeast}%
    \pgf@x=-1\pgf@x%
    \pgf@y=-.66666\pgf@y%
  }
  \anchor{CE}{
    \pgf@process{\northeast}%
    \pgf@x=-1\pgf@x%
    \pgf@y=-0.33333\pgf@y%
  }
  \anchor{Q}{
    \pgf@process{\northeast}%
    \pgf@y=.5\pgf@y%
  }
  \anchor{Qn}{
    \pgf@process{\northeast}%
    \pgf@y=-.5\pgf@y%
  }
  \anchor{R}{
    \pgf@process{\northeast}%
    \pgf@x=0pt%
  }
  \anchor{S}{
    \pgf@process{\northeast}%
    \pgf@x=0pt%
    \pgf@y=-\pgf@y%
  }
  \backgroundpath{
    \pgfpathrectanglecorners{\southwest}{\northeast}
    \pgf@anchor@dff@CLK
    \pgf@xa=\pgf@x \pgf@ya=\pgf@y
    \pgf@xb=\pgf@x \pgf@yb=\pgf@y
    \pgf@xc=\pgf@x \pgf@yc=\pgf@y
    \pgfmathsetlength\pgf@x{1ex} 
    \advance\pgf@ya by \pgf@x
    \advance\pgf@xb by \pgf@x
    \advance\pgf@yc by -\pgf@x
    \pgfpathmoveto{\pgfpoint{\pgf@xa}{\pgf@ya}}
    \pgfpathlineto{\pgfpoint{\pgf@xb}{\pgf@yb}}
    \pgfpathlineto{\pgfpoint{\pgf@xc}{\pgf@yc}}
    \pgfclosepath

    \begingroup
    \tikzset{flip flop/port labels} 
    \tikz@textfont

    \pgf@anchor@dff@D
    \pgftext[left,base,at={\pgfpoint{\pgf@x}{\pgf@y}},x=\pgfshapeinnerxsep]{\raisebox{-0.75ex}{D}}

    \pgf@anchor@dff@CE

    \pgf@anchor@dff@Q
    \pgftext[right,base,at={\pgfpoint{\pgf@x}{\pgf@y}},x=-\pgfshapeinnerxsep]{\raisebox{-.75ex}{Q}}

    \pgf@anchor@dff@Qn

    \pgf@anchor@dff@R

    \pgf@anchor@dff@S
    \endgroup
  }
}

\tikzset{ register/.pic={%
        \node[rectangle, minimum height=1.5cm, text width=0.8cm, text centered, draw] (-reg) {\scriptsize{\tikzpictext}};
        \node[draw, isosceles triangle, 
            isosceles triangle apex angle = 90, 
            minimum width = 0.25cm,inner sep=0pt,
            above right=0.2cm and 0cm of -reg.south west] (-clk) {};
        \node[below right= 0.1cm and 0cm of -reg.north west] (-D) {\scriptsize D};
        \node[below left= 0.1cm and 0cm of -reg.north east] (-Q) {\scriptsize Q};
        \coordinate (-out) at (-Q.east);
        
    }}

\tikzset{flip flop/port labels/.style={font=\ttfamily\scriptsize}}
\tikzset{add font/.code={\expandafter\def\expandafter\tikz@textfont\expandafter{\tikz@textfont#1}}} 
\tikzset{every dff node/.style={draw,minimum width=1cm,minimum 
		height=1.5cm,inner sep=1mm,outer sep=0pt,cap=round,add 
		font=\ttfamily}}
\tikzstyle{dataBus} = [draw, very thick]
\tikzstyle{fils} = [draw]

\makeatother

\tikzstyle{decision} = [diamond, draw, fill=blue!20, 
text width=4.5em, text centered, node distance=3cm, inner sep=0pt]
\tikzstyle{block} = [rectangle, draw, fill=blue!20, 
text width=7em, text centered, rounded corners, minimum height=3em]
\tikzstyle{line} = [draw, -latex']

\input{myTikzShape.tex}
\newsavebox{\tempbox}
\newsavebox{\tempboxi}

\usepackage[]{siunitx}
\DeclareSIUnit{\bit}{b}
\DeclareSIUnit{\nothing}{\relax}

\usepackage{array}
\usepackage{balance}
\usepackage{ifthen}

\begin{document}
%
\title{Bridging the Gap: FPGAs as Programmable Switches}
%
%
%
%
\author{Thomas~Luinaud\thanks{First three authors have equally contributed to this work.},
		Thibaut~Stimpfling,
		Jeferson~Santiago~da~Silva  \thanks{This research was partially funded by Mitacs/Canada and CPNq/Brazil.},\\
        Yvon~Savaria,
        and~J.M.~Pierre~Langlois,\\
        Polytechnique Montr\'{e}al, Canada\\
        \{firstname.lastname\}@polymtl.ca
    }
\IEEEoverridecommandlockouts

\maketitle

\begin{abstract}

  The emergence of P4, a domain specific language, coupled to PISA, a domain specific architecture, is revolutionizing the networking field. P4 allows to describe how packets are processed by a programmable data plane, spanning ASICs and CPUs, implementing PISA. Because the processing flexibility can be limited on ASICs, while the CPUs performance for networking tasks lag behind, recent works have proposed to implement PISA on FPGAs. However, little effort has been dedicated to analyze whether FPGAs are good candidates to implement PISA.

In this work, we take a step back and evaluate the micro-architecture efficiency of various PISA blocks.
We demonstrate, supported by a theoretical and experimental analysis, that the performance of a few PISA blocks is severely limited by the current FPGA architectures.
Specifically, we show that match tables and programmable packet schedulers represent the main performance bottlenecks for FPGA-based programmable switches.
Thus, we explore two avenues to alleviate these shortcomings.
First, we identify network applications well tailored to current FPGAs.
Second, to support a wider range of networking applications, we propose modifications to the FPGA architectures which can also be of interest out of the networking field.
\end{abstract}


\IEEEdisplaynontitleabstractindextext

%
\IEEEpeerreviewmaketitle

  \section{Introduction}\label{sec:intro}
\IEEEPARstart{T}{he} P4 language~\cite{Bosshart:2014:PPP:2656877.2656890} and the Protocol Independent Switch Architecture (PISA)~\cite{Bosshart:2013:FMF:2534169.2486011} have paved the road for high performance configurable data planes. P4 is a domain specific language (DSL) designed to describe how packets are forwarded on a data plane. PISA, a domain specific architecture associated to P4, provides a common abstraction to configurable data planes. Figure~\ref{fig:switch} presents PISA, a pipeline of configurable blocks (\S\ref{sec:back:PISA}).

However, current implementations of PISA switches limit the innovation potential of network architects.
Several network applications are difficult to implement in PISA switches, due to hardware constraints (stateful applications often require large shared embedded memory) or architectural rigidity (fixed-function packet scheduler).

Recently, FPGAs have emerged as a platform to accelerate network and server applications in data centers. 
As an example, Microsoft has deployed FPGAs in its data centers~\cite{cloud_scale_acceleration_architecture}.
In-network computing has also attracted attention of FPGA enthusiasts~\cite{ml_in_switch}.

Following this trend, recent research~\cite{p4fpga,p4-netfpga} has also proposed mapping PISA to FPGAs. 
The goal is to exploit the inherent FPGA reconfigurability to implement network applications as new P4 programs can be deployed by simply reconfiguring the FPGA bitstream. 
Yet, the performance of FPGA-based PISA switches are, at best, one order of magnitude lower than that of their ASIC counterparts. 

In this work, we take a step back and we thoroughly analyze \textit{how} PISA blocks are implemented in FPGAs to identify the strengths and weaknesses of such mapping.
These theoretical and experimental analyses have led to two major conclusions. 
First, some network applications are intrinsically good matches for FPGA implementation. 
Second, FPGA devices need to be reengineered to better support networking applications.

In summary, our contributions are as follows:
\begin{enumerate}[noitemsep]
	\item We analyze the mapping of PISA blocks to FPGAs, highlighting the pros and cons~(\S\ref{sec:prog_dp});
	\item We evaluate, analytically and experimentally, the performance of FPGA-based PISA switches~(\S\ref{sec:scaling_throughput});
	\item We identify which applications are well suited for current FPGAs~(\S\ref{sec:shine}); and
	\item We propose the specialization of the FPGA architecture for the network domain~(\S\ref{sec:specialized_fpga}).
   \end{enumerate}


\section{Background}\label{sec:back}

A generic FPGA architecture is first presented. Then, the configurable blocks of PISA are introduced.

\subsection{FPGA architecture}\label{sec:back:fpga}
\begin{figure}[t]
    \centering
\begin{tikzpicture}[node distance=0.75cm, every node/.style={draw}]
\tikzstyle{bram} = [rectangle, draw, , minimum width=.30cm, minimum height=0.6cm,inner sep=0]
\tikzstyle{dsp} = [circle, draw, minimum width=0.25cm, inner sep=0]
\tikzstyle{slice} =[rectangle, draw, minimum width=.2cm, minimum height=0.1cm,inner sep=0]
\tikzstyle{IOBlock} =[rectangle, draw, minimum width=.1cm, minimum height=.1cm,inner sep=0]
\tikzstyle{XYBlock} =[rectangle, draw, minimum width=.5cm, minimum height=.5cm,inner sep=0]
\pgfmathsetmacro{\XYBlockGrid}{3}
\pgfmathsetmacro{\nbIO}{12}
\pgfmathsetmacro{\gridspace}{0.135}
\begin{scope}
	\coordinate (L0) at (0,0);
	\node[XYBlock,anchor=north west] (BB00) at (L0) {};
	\foreach \y [evaluate=\y as \oy using int(\y - 1)] in {1,...,\XYBlockGrid}{
		\node[XYBlock, below=\gridspace of BB\oy0](BB\y0){};
	}
	\foreach \x [evaluate=\x as \ox using int(\x - 1)] in {1,...,\XYBlockGrid}{
		\node[XYBlock, right=\gridspace of BB0\ox](BB0\x){};
		\foreach \y [evaluate=\y as \oy using int(\y - 1)] in {1,...,\XYBlockGrid}{
			\node[XYBlock, below=\gridspace of BB\oy\x](BB\y\x){};
		}
	}
	\foreach \x [evaluate=\x as \ox using int(\x - 1)] in {1,...,\XYBlockGrid}{
		\path[draw, very thick] ($(BB\ox0.south west) - (0,\gridspace/2)$) --
						   ($(BB\ox\XYBlockGrid.south east) - (0,\gridspace/2)$)
						   ($(BB0\ox.north east) + (\gridspace/2,0)$) -- 
						   ($(BB\XYBlockGrid\ox.south east)  + (\gridspace/2,0)$);
	}
	\coordinate (interconnect) at ($(BB00.south east) + (\gridspace/2,-\gridspace/2)$);
	
	\coordinate (MXMY) at (BB\XYBlockGrid\XYBlockGrid.south east);
	\coordinate (IO00) at ($(BB00.north west) + (-0.1, 0.1)$);
	\coordinate (IO10) at ($(BB00.north west) + (-0.1, 0.1)$);
	\coordinate (IOe00) at ($(MXMY) + (0.1,-0.1)$);
	\coordinate (IOe10) at ($(MXMY) + (0.1,-0.1)$);
	\foreach \x [evaluate=\x as \ox using int(\x - 1)] in {1,...,\nbIO}{	
		\node[IOBlock, below=0.1cm of IO0\ox](IO0\x){};
		\node[IOBlock, right=0.1cm of IO1\ox](IO1\x){};
		\node[IOBlock, above=0.1cm of IOe0\ox](IOe0\x){};
		\node[IOBlock, left=0.1cm of IOe1\ox](IOe1\x){};
	}
\end{scope}

\node[circle, minimum height=0.45cm, inner sep=0] (InterNote) at (interconnect) {};
\coordinate (zoomInterOri) at (IO01.west |- interconnect);
\node[anchor=east, minimum height=0.5cm, minimum width=1cm, densely dash dot] (zoomInter) at ($(zoomInterOri) - (0.3,0)$) {};
\path[draw, densely dashed] (InterNote.north west) -- (zoomInter.north east)
							(InterNote.south west) -- (zoomInter.south east);
							
\node[minimum height=0.3cm, minimum width=0.5cm] (interconnectSym) at (zoomInter) {};
\pgfmathsetmacro{\switchSpaceY}{0.05}
\pgfmathsetmacro{\switchSpaceX}{0.15}
\path[stealth-stealth, draw] ($(interconnectSym.north west) - (\switchSpaceX, \switchSpaceY)$) -- ($(interconnectSym.north west) - (-\switchSpaceX, \switchSpaceY)$) --
							 ($(interconnectSym.south east) - (\switchSpaceX, -\switchSpaceY)$) -- ($(interconnectSym.south east) + (\switchSpaceX, \switchSpaceY)$);
\path[stealth-stealth, draw] ($(interconnectSym.south west) - (\switchSpaceX, -\switchSpaceY)$) -- ($(interconnectSym.south west) + (\switchSpaceX, \switchSpaceY)$) --
							 ($(interconnectSym.north east) - (\switchSpaceX, \switchSpaceY)$) -- ($(interconnectSym.north east) + (\switchSpaceX, -\switchSpaceY)$);

\begin{scope}[node distance=0.2cm]
	\node[slice, right = 1cm of IOe011] (s00) {};
	\foreach \x [evaluate=\x as \ox using int(\x - 1)] in {1,...,3}{	
		\node[slice, below=0.05cm of s0\ox](s0\x){};
	}
	\node[dsp, right=0.3 of s00.north east, anchor=north] (d00) {\tiny $\times$};
	\node[dsp, below=0.1 of d00] (d01) {\tiny $\times$};
	\node[bram, right=0.3 of d00.north, anchor=north west] (b00) {};
	
	\node[slice, below = of s03] (s10) {};
	\foreach \x [evaluate=\x as \ox using int(\x - 1)] in {1,...,3}{	
		\node[slice, below=0.05cm of s1\ox](s1\x){};
	}
	\node[dsp, right=0.3 of s10.north east, anchor=north] (d10) {\tiny $\times$};
	\node[dsp, below=0.1 of d10] (d11) {\tiny $\times$};
	\node[bram, right=0.3 of d10.north, anchor=north west] (b10) {};
	\node[fit = (s00) (b10), densely dash dot] (BBzoom) {};
\end{scope}

\path[draw, densely dashed] (BB13.north east) -- (BBzoom.north west)
			(BB13.south east) -- (BBzoom.south west);
\begin{scope}[every node/.style={draw=none, font=\small, inner sep=0.2pt}]
	\node[above=0.2cm of zoomInter, draw=none] (lInter) {Crossbar};
	\path[draw] (zoomInter) -- (lInter);
	\node[draw=none, anchor=east, text width=1.1cm] (lFabric) at ($(BB10.south -| zoomInter.east) - (0,\gridspace/2)$)  {Routing Fabric};
	\path[draw] (lFabric.east) -- ($(BB10.south west) - (0,\gridspace/2)$);

	\node[anchor=south west] (lDSP) at ($(d00.north -| BBzoom.east) + (0.2,0.1)$) {DSP};
	\node[anchor=west] (lBRAM) at (b10.east -| lDSP.west) {BRAM};
	\node[anchor=north west] (lSlice) at ($(s13.south -| lDSP.west) + (0,-0.1)$) {Slice};
	\node[anchor=west] (lIO) at (IOe01.east -| lDSP.west) {I/O};
\end{scope}

\path[draw] (s13) -- ($(s13.south |- lSlice.west) + (0.4,0)$) -- (lSlice);
\path[draw] (d00.north) -- ($(d00.north |- lDSP.west) + (0.4,0)$) -- (lDSP);
\path[draw] (b10.east) -- (lBRAM);
\path[draw] (IOe01.east) -- (lIO);
\end{tikzpicture}
    \caption{Considered FPGA architecture}
    \label{fig:fpga}
\end{figure}
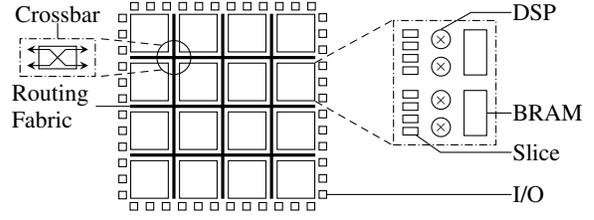

As shown in Figure~\ref{fig:fpga}, FPGAs are structured as an array of blocks interconnected by a routing fabric. The routing fabric allows to route signals between the FPGA blocks.

To communicate off-chip, FPGAs integrate configurable Input/Output (I/O) blocks and dedicated high-speed transceiver pins. Hard-wired PCIe blocks and MAC blocks, connected to the high-speed transceivers pins, are also integrated in modern FPGAs.

Logic operations are implemented by Configurable Logic Blocks (CLBs), distributed over slices. A slice comprises lookup tables (LUTs), which implement logic functions, and flip-flops (FFs) used to synchronize signals between logic functions.

In addition, specialized blocks are also hard-wired into FPGAs, such as Digital Signal Processing (DSP) blocks and Block RAM (BRAM). DSPs perform arithmetical operations, while BRAMs are width-configurable SRAMs. Hard-wired blocks avoid wasting CLBs and allow higher clock frequencies.

\subsection{Protocol Independent Switch Architecture}\label{sec:back:PISA}

Historically, network switches have been build upon fixed function ASICs as packets undergo similar and straightforward processing in switch devices.
However, the advent of Software-Defined Networking (SDN) has changed this state of affairs as it mandates for network programmability.

\begin{figure}[]
  \centering
\begin{tikzpicture}[node distance=0.75cm, every node/.style={draw}]
\tikzstyle{match} = [rectangle, draw, fill=blue!20, minimum width=.65cm, minimum height=0.5cm]
\tikzstyle{action} = [shape=trapezium, draw, trapezium angle=-70,fill=yellow!20, minimum width=0.5cm, shape border rotate=90]

\node[draw=none] (s0){};
	\node[match, below right=0.25cm and 0.5cm of s0.north east,anchor=north west](m0){};
	\node[action, right=0.2cm of m0](a0){};
	\foreach \x [evaluate=\x as \ox using int(\x - 1)] in {1,...,3}{	
		\node[match, below=0.1cm of m\ox](m\x){};
		\node[action, right=0.2cm of m\x](a\x){};
	}
	\node[fit=(m0) (a0) (m3) (a3), inner sep=0.15cm] (s1) {};

\node[minimum height=2.65cm, minimum width=.85cm, left=0.5cm of s1.west, anchor= east] (parser) {};
\node[text centered, inner sep = 2pt, text width=3.0cm-8pt, rotate=90, draw=none] at (parser) {\small Programmable Parser};

\node[minimum height=2.65cm, minimum width=0.85cm, right= 0.5 of s1] (Scheduler) {};
\node[text centered, inner sep = 2pt, text width=2.0cm-8pt, draw=none, rotate=90] at (Scheduler) {\small Packet Scheduler};

	\node[match, below right=0.15cm and 0.65cm of Scheduler.north east,anchor=north west](m0){};
	\node[action, right=0.2cm of m0](a0){};
	\foreach \x [evaluate=\x as \ox using int(\x - 1)] in {1,...,3}{	
		\node[match, below=0.1cm of m\ox](m\x){};
		\node[action, right=0.2cm of m\x](a\x){};
	}
	\node[fit=(m0) (a0) (m3) (a3), inner sep=0.15cm ] (s2) {};

\node[minimum height=2.65cm, minimum width=0.85cm, right= 0.5 of s2] (deParser) {};
\node[rotate=90, draw=none] at (deParser) {\small Deparser};

\coordinate (input) at ($(parser.north west) + (-.3cm, 0)$); 
\coordinate (output) at ($(deParser.north east) + (.38cm, 0)$);

	\foreach \y in {0.25,0.69,...,2.65}{
		\coordinate (temp) at ($(parser.south west) + (0,\y)$);
		\path[thick, -Computer Modern Rightarrow] (input|-temp) edge (parser.west|-temp);
		\path[thick, -Computer Modern Rightarrow] (deParser.east|-temp) edge (output|-temp);
	}

\path[line width=1.5pt, -Stealth]
(parser) edge (s1) 
(s1) edge (Scheduler)
(Scheduler) edge (s2)
(s2) edge (deParser);

\node[fit=(parser) (s1) (Scheduler) (s2) (deParser), inner xsep=0.2cm, inner ysep=0.1cm] (s2) {};
\end{tikzpicture}
  \caption{Reference PISA switch model}
  \label{fig:switch}
\end{figure}
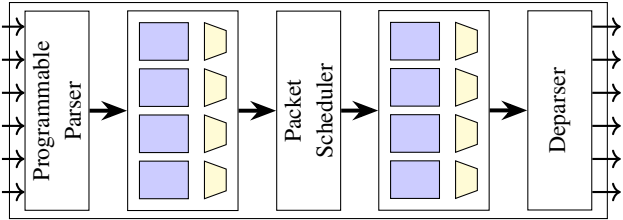

In this context, the PISA architecture~\cite{Bosshart:2013:FMF:2534169.2486011} was the first proposal to support programmable protocol-agnostic packet forwarding. 
PISA comprises a programmable \textit{parser}, a programmable pipeline of \textit{match-action tables}, a fixed function \textit{packet scheduler}, and a configurable \textit{deparser}.
PISA is illustrated in Figure~\ref{fig:switch}, where blue rectangles are match tables and yellow trapeziums are Arithmetic and Logical Units (ALUs) implementing actions.

A parser extracts header fields that are used to build lookup keys for the match-action stages. A key is matched against rules stored in a match table. The lookup result, an action and an associated data, is executed by an action stage. A packet scheduler reorders the packet according to a scheduling algorithm. Finally, a deparser reassembles the updated packet headers and emits the packet.

\section{Mapping PISA to FPGAs}\label{sec:prog_dp}
A typical implementation of PISA on FPGA follows a dataflow architecture, where the processing is laid out \textit{spatially} on a pipeline.
That is, the processing is divided into a sequence of operations, where each operation is mapped onto a portion of the FPGA.
Packets are streamed on a data bus throughout the PISA components. Because a packet size can be larger than the bus width, a packet can be segmented over the data bus, and can require multiple clock cycles to traverse a PISA component.

In the next sections, we describe the key components of the PISA architecture and we discuss their micro-architecture efficiency when implemented on FPGAs.

\subsection{Programmable Parser}\label{sec:prog_dp:parser}

The packet parser extracts header fields to be matched in the match-action stages, and determines the stack of valid protocols in a switch~\cite{Gibb:2013:DPP:2537857.2537860}.
The common approach to implement packet parsers is through an abstract state machine.
The state machine can be represented as a Directed Acyclic Graph (DAG) in which nodes are protocols and edges indicate protocol transitions.

DAGs are a good fit to FPGAs due to their embarrassingly pipeline-able characteristic.
Also, data flow architectures can be leveraged by the intrinsic of the FPGA fabric, where logic elements are tightly coupled to registers improving the pipeline efficiency.
In addition, nodes of a packet parser DAG are compact data structures that easily fit in on-chip memories or registers.
As a result, multiple high-performance programmable packet parsers have been implemented on FPGAs~\cite{Attig:2011:GPP:2065093.2065215,BENACEK201822,SantiagodaSilva:2018:PHS:3174243.3174270}.




\subsection{Match Stage}\label{sec:prog_dp:ma}\label{sec:prog_dp:ma:match}
A match table is an abstract container, holding a collection of keys, actions and data, addressed by a lookup key.
In P4, a match table can be configured for exact match, ternary match or longest prefix match.

\textbf{Exact Match (EM).}
EM is traditionally implemented with content addressable memory (CAM).
A CAM memory is associative array used to find if an exact key value is stored into a table. 
Such CAM memories are no longer integrated as hard blocks in FPGAs and must be emulated.

One approach to CAM emulation is \textit{transposed} memories, where the lookup key is used as an address. The memory stores for each lookup key a bitmap of the matched keys. The \textit{transposed} approach typically yields a memory efficiency less than 10\%~\cite{gap_fpga_cmos,cam_ubc}.

An efficient approach to implement exact match operations exploits a hash table combined with Cuckoo hashing for collision resolution, yielding a memory efficiency higher than 80\%~\cite{Bosshart:2013:FMF:2534169.2486011,power-one-move}. Thus, EM implementation on FPGAs can be efficient using on-chip memories.


\textbf{Ternary Match.}\label{sec:tcam}
TCAM are traditionally used for ternary matches. 
Hardware TCAM comprises a memory with a match circuitry, to store a ternary rule and match it against a lookup key and a priority encoder, which returns the matched rule index with the highest priority.
Similarly to CAMs, TCAMs are emulated on FPGAs.

A memory efficient approach to emulate a TCAM of width $W$ and depth $N$ is to build $P$ smaller TCAMs using the transposed memory approach.
Still, the memory overhead is $2^{w}/w$, with $w=W/P$.
Hence, the overhead is minimal for a TCAM where $w=1$ or $w=2$, which translates to a depth of 2 or 4.
In practice, this overhead ranges from $8.4\times$ to $65\times$, since current FPGA memories have a minimum depth of 32 (LUT RAMs) and 512 (BRAMs).

\citeauthor{pr_tcam} propose to represent ternary rules as logic functions~\cite{pr_tcam}, which are synthesized in LUTs~(\S~\ref{sec:back:fpga}).
However, this approach yields a memory efficiency similar with the transposed memory approach, because a LUT is a small SRAM that records one result for each possible input. Hence, implementation of efficient TCAMs on FPGAs remains an open question.

\textbf{Longest Prefix Match (LPM).}
LPM is a sub-case of ternary match with two differences: 1) the mask applies on a contiguous segment of bits, starting from the least significant bit, 2) when multiple objects match a lookup key, only the longest prefix matched is returned.
Thus, LPM implemented with TCAMs~\cite{lpm_ubc} are achieving a low memory efficiency on FPGAs (\S~\ref{sec:tcam}).

Otherwise, LPM can be emulated with data structures such as binary trees as in the Xilinx LPM IP. Not only both the memory efficiency and frequency are improved by 2$\times$, but the resource consumption is reduced by an order of magnitude over transposed memory. However, the update latency grows linearly with the number of keys.

Several other data structures proposed in the literature~\cite{2-3-tree, fib_compression_entropy}, exploit specific characteristics of the stored key to improve memory efficiency.
However, these methods can hardly be used in systems configured with P4 since the characteristics of the content of match tables are unknown a priori.

\subsection{Action Stage}\label{sec:prog_dp:ma:actions}

\textit{Primitive actions} are operations natively supported in P4, comprising arithmetic, logic, bit shifts, and conditional operations.

\textbf{Arithmetic and logic operations.} These operations are, except for the division, efficiently mapped to the FPGA fabric.
Indeed the ALU found in a CPU is efficiently implemented on FPGA~\cite{gap_fpga_cmos}.
In addition, modern FPGAs supports fast multiplications and multiply-and-accumulate operations, as well as adders with DSPs and hard-wired carry chains. However, division operations on FPGAs are costly in terms of logic resources, but are quite uncommon in packet processing.
Complex actions combining multiple primitives are well supported when laid out in a pipeline.

In addition, the reconfiguration capability of FPGAs allow supporting almost any combination of actions.

\textbf{Conditional operations.} Conditions described in P4 are translated at the architectural level into multiplexers that select one operation, or result, out of multiple inputs.
While an FPGA slice can be configured as a 16 to 1 multiplexer, very wide multiplexers must be distributed over multiple slices, which degrades performance with the number of conditions.

\textbf{Bit-shift operations.}
Common fixed bit shifts represent no hardware cost when implemented in FPGA because the shift values are known at compile time, hence, they are hardwired into the FPGA fabric.
Likewise, barrel shifters are poorly mapped in FPGAs as they are commonly implemented using a chain a muxes. These, in turn, are uncommon operations in packet processing, therefore, when used, barrel shifters can be implemented using hard DSP block.



\subsection{Packet Scheduler}\label{sec:prog_dp:scheduler}
A packet scheduler decides at what times and in what orders are packets sent.

The push-in-first-out (PIFO) queue~\cite{Sivaraman:2016:PPS:2934872.2934899} was recently proposed as an abstraction upon which a programmable packet schedulers can be built. However the PIFO architecture maps poorly to an FPGA, because a range-search CAM is required, which can only be emulated either with flip-flops, or with the transposed memory approach (\S\ref{sec:prog_dp:ma:match}).

Alternatively, \citeauthor{Benacer:18} have proposed a priority queue that better fits to the FPGA architecture~\cite{Benacer:18}.
It exploits the intrinsic parallelism of FPGAs to sort the packet ranks using a systolic priority queue. However, this work does not provide the programmability supported by the PIFO. 

In addition, a packet buffer is required to store packets while they are scheduled.
Assuming a typical data center \SI{100}{\micro\second} RTT and one \SI{100}{\giga\bit/\second} interface, an \SI{1.2}{\mega\nothing}B buffer is required. Hence, assuming an FPGA with 12$\times$\SI{100}{\giga\bit/\second} interfaces, a \SI{12}{\mega\nothing}B buffer would be required, which would use one fourth of the on-chip memories available on the largest FPGAs and would stress the FPGA internal routing fabric (\S\ref{sec:scaling_throughput}).



%
%

\subsection{Deparser}\label{sec:prog_dp:deparser}
The deparser is a module that performs the inverse of what the parser does as it reassembles headers in the correct order before sending the packet.
Thus, a deparser is well supported on FPGAs, and the methods used to implement a high performance parser can be applied~\cite{BENACEK201822}. 

In P4\textsubscript{16}, the deparser is described as a sequence of header emission statements.
The packet header is then recomposed by respecting the order of valid headers in the sequence.
However, the headers validity can be modified during the match-action stages and current compilers do not lifetime header analysis to find the smallest combination set of possible valid headers.
Thus, the deparser is more difficult to implement and it tends to have the largest resource consumption in the P4 pipeline~(\S~\ref{sec:experimental}).

\section{Scaling the Packet Throughput}\label{sec:scaling_throughput}
High-end FPGAs come with multiple hard-wired \SI{100}{\giga\nothing} Ethernet MACs, offering a total packet throughput at the I/O level exceeding the \SI{}{\tera\bit/\second} barrier. This raises the question whether an FPGA  can \textit{in practice} process packets at a \SI{}{\tera\bit/\second}.

Because PISA is implemented on FPGAs using a data flow architecture, the packet throughput supported is directly the product $width_{bus} \times frequency_{bus}$. Hence, increasing either the bus width, or the bus frequency directly translates to a higher packet throughput. Another option to scale the packet throughput is to use parallelism, i.e, to replicate a PISA pipeline. We first discuss each of the three approaches. Second, based on experimentation, we present \textit{in practice} the scaling limitations on FPGAs. 

\subsection{Methods}
\textbf{Scaling the Bus Width.}
To simplify the discussion, we assume that the minimum packet size is greater or equal to the bus width. Increasing the bus size comes at the cost of a higher resource consumption and limits the maximum bus frequency. Because the bus size is increased, more bits are synchronized and processed, which directly increases the resource consumption. However, wide buses ($>$512 bits) increase the routing congestion in FPGAs, leading to longer wire delays, which directly limits the frequency.

\textbf{Scaling the Bus Frequency.} One method consists in increasing the depth of a pipeline to reduce the logic delay and wire delay between two flip-fops.
The latency can increase, but the shorter clock periods can be obtained, which increases the frequency.
In theory, the parser, match action tables and deparser can be heavily pipelined, and thus, the frequency can scale. However, experimentally, the FPGA architecture limits the frequency scaling~(\S~\ref{sec:experimental}). 

\textbf{Pipeline Replication.} 
Our experiments~(\S~\ref{sec:experimental})~show that the maximum practical throughput in single pipeline is around \SI{800}{\giga\bit/\second}. 
However, State-of-the-art FPGAs can support almost twice this throughput as hard \SI{100}{\giga\bit/\second} MAC blocks.

The packet throughput can be increased linearly with the number of PISA pipelines implemented, at the cost of a linear resource consumption growth.

To exploit the benefits of pipeline replication, a packet dispatcher is required at both the input and output of the pipelines to distribute the packet traffic among the pipelines. In addition, the packet dispatcher integrates buffers in each port to prevent packet drops. The resulting buffers complexity can be expressed as:

\begin{align*}
  \mathtt{BufferSize = \frac{Ports}{Pipes} \times MaxPktSize}\\
  \mathtt{TotalBuffer Size = 2 \times Ports \times BufferSize}
\end{align*}
where $\mathtt{Pipes}$ is the number of pipelines, $\mathtt{Ports}$ the number of Ethernet ports, $\mathtt{BufferSize}$ the input/output buffer size per port, $\mathtt{MaxPktSize}$ the maximum supported packet size, and $\mathtt{TotalBufferSize}$ the total required buffer size.


\begin{figure}[t]
\centering
\includegraphics[width=0.5\textwidth]{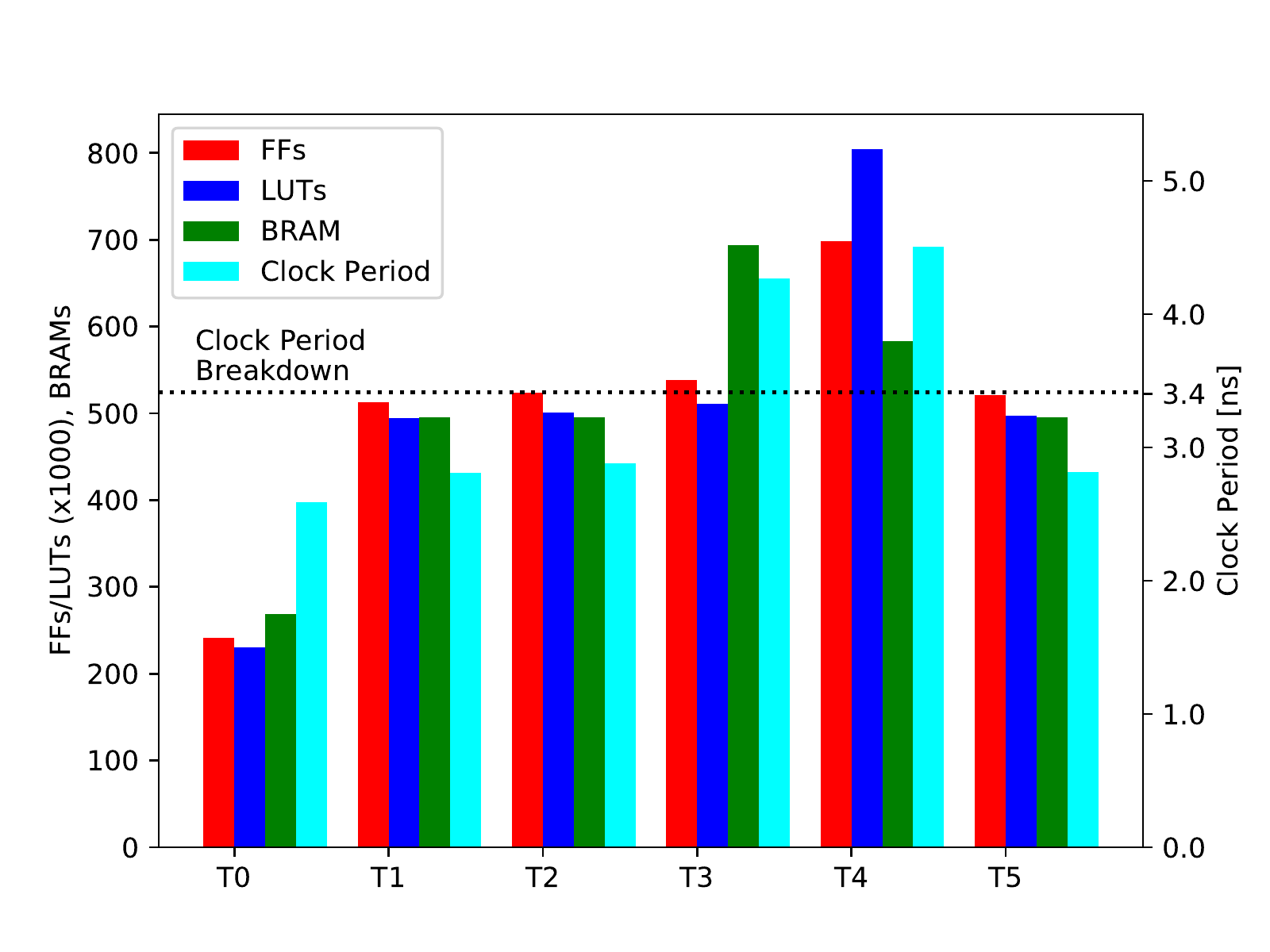}
\caption{FPGA results for test cases from Table~\ref{tab:test_case}.}
\label{fig:test_cases}
\end{figure}

\subsection{Experimental Evaluation}\label{sec:experimental}
To characterize some FPGA performance limitations, we first evaluate the impact of the data bus width on the clock frequency. Then, we characterize the performance of multiple PISA blocks in terms of resource usage and clock frequencies. The PISA block stressed and the description of each test is presented in Table~\ref{tab:test_case}. All the experiments were described in P4\footnote{The source code is available at \url{https://github.com/luinaudt/Unleashing_FPGA}}.
The implementation was executed with Xilinx SDNet 2017.4 combined with Vivado 2018.2 on a Xilinx Virtex Ultrascale+ FPGA (XCVU9P-flga2577-3-e) with a targeted clock frequency of \SI{500}{\mega\hertz}.

\textbf{Bus width vs frequency.} 
To characterize the bus width impact on the clock period, the test T0 was implemented with a bus width ranging from 64 to 2048 bits. 

\begin{figure}[t]
\centering
\includegraphics[width=0.5\textwidth]{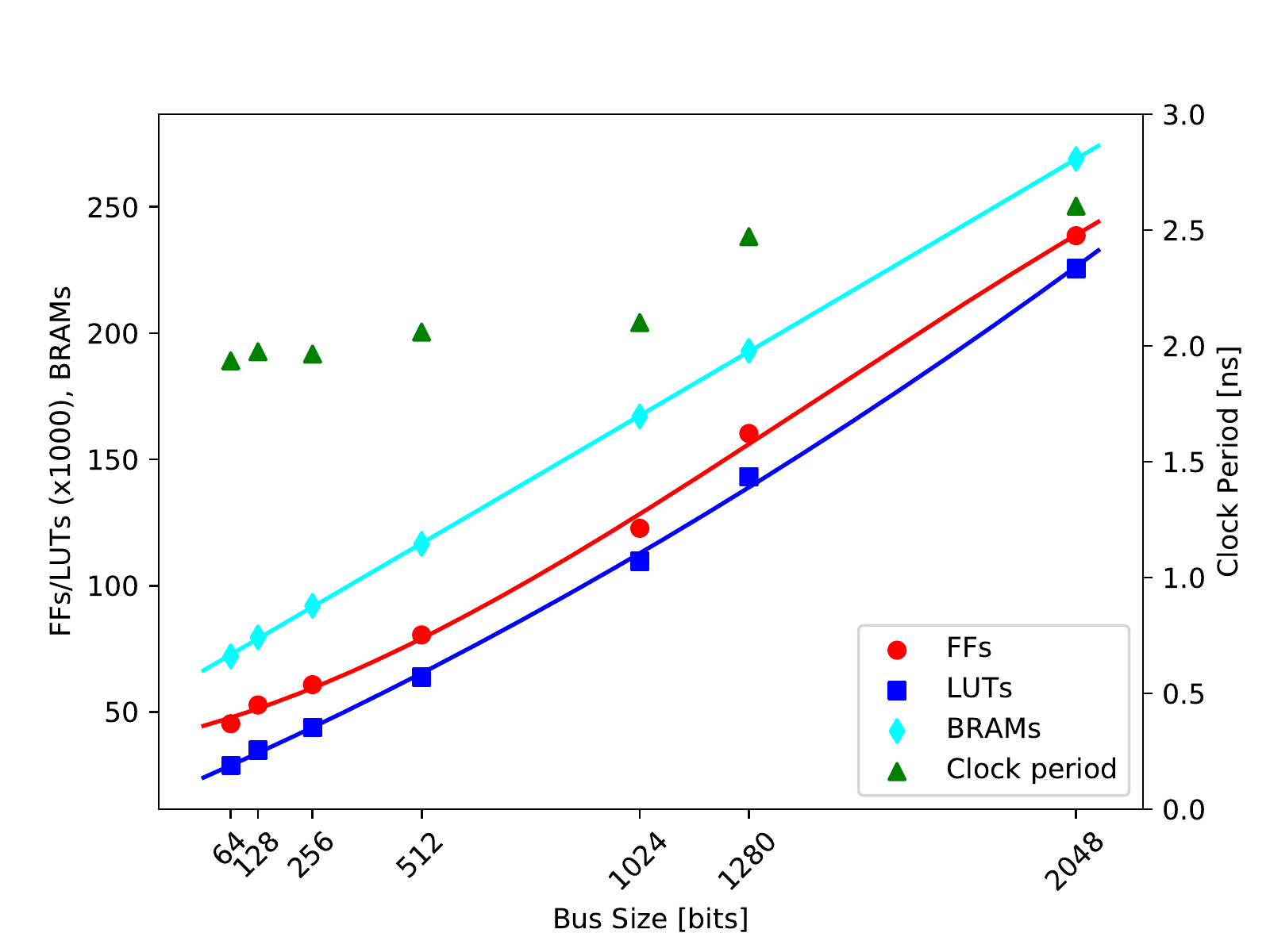}
\caption{Resource consumption and clock period as function of the bus size for the T0 benchmark.}
\label{fig:bus_size}
\end{figure}

As shown in Figure~\ref{fig:bus_size}, the resource consumption is linear with the bus width, while the maximum clock frequency decreases for bus larger than 1280 bits. This frequency deterioration relates to routing congestion and a large net fan-out. When a routing block is congested, signals are routed to farther routing blocks, which increases wire delays. In addition, we observe a large fan out in our tests that limit the maximum clock frequency, even with hyper-pipelined data paths.

We observe a maximum throughput around \SI{786}{\giga\bit/\second} per implemented pipeline for a bus width of 2048 bits. As a result, two pipelines are required to process the I/O bandwidth available on high-end FPGAs.


%
%
%

\textbf{Stressing PISA.}
To stress the different PISA blocks, we used a bus width of 2048 bits. The implementation results are presented in Figure~\ref{fig:test_cases}. The dotted line shows the clock period required for a throughput of \SI{600}{\giga\bit/\second} per pipeline.

We observe that T1, T2 and T5 have little impact on the clock period. Thus, increasing the protocol stack and adding more action does not affect throughput. By contrast, in T3 and in T4, the clock period is increased by more than 60\%. The performance deterioration for T3 does not relate to the CAM emulation method \S~\ref{sec:prog_dp:ma:match}, but to the BRAMs distribution over the FPGA fabric. To implement larger memories, several BRAM blocks are combined. Because, BRAMs are organized in columns~(\S~\ref{sec:back:fpga}), large memories are spread into multiple BRAM columns, incurring long wire delays, which decrease the clock period. For T4, the reduced performance also relates to the use of large distributed RAM spanning over the FPGA fabric. 

Finally, resource consumption increases between T0 and T1, because the deparser uses more than 80\% of all resources.
The LUT and FF usage is similar for T1, T2, T3 and T5, which demonstrates the efficiency of actions performed on FPGAs. Compared to T1, more BRAMs are used for T3 and T4 to implement the match tables. T4 uses a high number of LUTs and FF, because of an inefficient TCAM implementation.

\begin{table}[t]
	\centering
	\caption{Test cases to stress the FPGA architecture.}
	\label{tab:test_case}
	\begin{tabular}{lm{4.5cm}m{2.0cm}}
		\toprule
		{Test}      & {Description}              & {Stress}          \\
		\midrule
		T0          & 3 headers, 74 bytes, no tables                   & -                 \\
		T1          & T0 + 8 headers, 116 bytes     & (De)Parser  \\ 
		T2          & T1 + IPv4 checksum    & Actions           \\ 
		T3          & T2 + EM \SI{64}{\kilo\nothing}~$\times$~128 bits            & Match tables      \\
		T4          & T2 + TCAM \SI{4}{\kilo\nothing}~$\times$~128 bits            & Match tables      \\
		T5          & T2 + 4 chained conditionals      & Actions           \\
		\bottomrule
		\hline
	\end{tabular}
\end{table}

\section{FPGAs and In-Network computing}\label{sec:shine}
A growing interest in the literature is shown for in-network computing. Our thesis is that FPGAs are a key component to unlock the potential of this paradigm. Indeed, FPGAs can support a very high packet throughput in several cases~(\S~\ref{sec:experimental}).
This raises three fundamental questions:
\begin{enumerate}[noitemsep]
\item Which applications can efficiently be implemented in current  FPGA?
\item Which programming model to use? 
\item Is PISA sufficient for in-network computing?
\end{enumerate}

\textbf{Applications tailored to FPGAs.} 
\citeauthor{when-should-the-network-be-the-computer} classify in-network computing applications in three categories: the number of operations per packet, the number of states per packet and the packet gain~\cite{when-should-the-network-be-the-computer}. Based on our experimentations, applications tailored to FPGAs need to perform a high number of operations per packet, have a limited number of states, but avoid match tables.

An example of application meeting these criteria is the reduction operation in a distributed deep neural network (DDNN) training~\cite{ml_in_switch}. The DDNN training consists of computing gradients which are subsequently collected by a parameter server (PS). The PS aggregates received values with arithmetical operations, then updates the model and forward it back.
Because the PS has a high number of operations per packet, packets are mainly parsed and forwarded, this application could support a higher packet throughput over the previously reported results~\cite{ml_in_switch}. 

\textbf{Programming model.} To enforce the abstraction layers, we argue that a network DSL, such as P4, should be used only to describe packet manipulation, while the remaining application processing should be expressed with a general-purpose language.
In addition, current high level synthesis tools, synthesizing a general-purpose language for FPGAs, have shown to achieve performance close to hand written RTL code~\cite{SantiagodaSilva:2018:PHS:3174243.3174270}.

\textbf{Abusing PISA externs.} On FPGAs, an in-network computing application can consist of PISA connected to an application module. P4 \textit{externs} allow to connect external hardware modules to PISA. Hence a generic in-network computing architecture would integrate standardized interfaces to extern processing modules.

\section{Specializing FPGAs}\label{sec:specialized_fpga}
PISA efficiency and performance are currently limited by the existing FPGA architecture. Notably, the match stages and packet scheduler are the main performance bottlenecks.

We propose to specialize the FPGA architecture to better support PISA, while preserving the FPGA's flexibility for applications outside of the networking realm.


\textbf{Hard-wired TCAMs.} This would increase resource efficiency and performance of a ternary match stage.


To evaluate the benefits of hard-wired TCAMs, we analyze the number of transistors required for a 48$\times$128 TCAM as older Lattice FPGAs had hard-wired TCAMs using this configuration.
It is well established that a single-bit TCAM requires 16 transistors.
Thus, \SI{98}{\kilo\nothing} transistors are needed for the memory and match circuitry in a 48$\times$128 TCAM block.
The cost associated to the priority encoder is ignored as it is negligible.

In contrast, a 1-bit SRAM cell requires 6 transistors. Since soft-TCAMs use SRAMs, and have a 10$\times$ memory overhead (\S~\ref{sec:prog_dp:ma:match}), a 48$\times$128 soft-TCAM needs \SI{368}{\kilo\nothing} transistors. 
Hence, ignoring the reconfigurability cost, a hard-wired TCAM could reduce the silicon usage by 3.8$\times$.




Determining the number of TCAM primitives to hard-wire on FPGA is left for future work, as it relates to the FPGA thermal budget and the TCAM cells used~\cite{Bosshart:2013:FMF:2534169.2486011}.


More flexible on-chip memory primitives can be as an alternative to hard-wired TCAMs.
The transposed memory approach would directly benefit from shallower and wider memories.
However, wider memories imply large data buses, which have a direct impact on performance
~(\S\ref{sec:scaling_throughput}).

\textbf{Hard-wired CAM.} The motivation to hard-wire CAMs is to better support packet schedulers, but \textit{not} for lookup operations.
Indeed, the PIFO micro-architecture~\cite{Sivaraman:2016:PPS:2934872.2934899} requires single cycle CAMs, which are poorly emulated on FPGAs~\cite{cam_ubc}. 
In addition, PIFO uses range-search CAMs, where a lookup key must be enclosed in a range.
However, a hard range-search CAM is too specialized to be included as a generic module. 

Hence, we propose to hard-wire configurable CAM primitives supporting $=$, $<$ and $>$ match operations, which brings two benefits. 
First, multiple CAM primitives can be combined with programmable logic to construct range search CAMs. Second, because multiple match operations are supported, other applications can use it. For instance, soft-CPUs can use the proposed CAM for cache units, out-of-order scheduling~\cite{}, and applications using sparse matrices, such as machine learning, to reduce memory footprint.



We now evaluate the CAM size needed for the PIFO architecture.
The CAM size derives from the number of flows and the rank size supported by a packet scheduler.
To support 1024 flows, with a rank size of 16 bits, a range-search CAM of 1024 $\times$ (16 $\times$ 2) is required~\cite{Sivaraman:2016:PPS:2934872.2934899}. 
\citeauthor{Sivaraman:2015:TPP:2834050.2834106}~\cite{Sivaraman:2015:TPP:2834050.2834106} report an overhead of almost 20$\times$ between a range-search CAM block and its equivalent size SRAM.
However, since the analyzed CAMs are built using flip-flops, this overhead is over-evaluated by a factor of 3.
In addition, because high-end FPGAs pack hundreds of \SI{}{\mega\bit} of SRAM, using hundreds of \SI{}{\kilo\bit} of SRAM for CAMs has a very limited impact. 


\textbf{Dedicated Network on Chip.} One solution to limit the signal congestion inside the routing fabric is to integrate a Network on Chip (NoC). A NoC can be seen as a grid interconnecting different blocks of the FPGA together.
Because a NoC is hard-wired, it supports high clock frequency, which in turns allows a high packet throughput as presented by Achronix\footnote{https://www.achronix.com/product/speedster7t/}.
Also, a NoC simplify routing complexity by reducing the routing scope.

\textbf{Hard-wired wide buses.} We propose to integrate hard-wired wide buses into the routing fabric in order to support a high packet throughput. The idea consists in routing a n-bit bus as a single instance instead of routing individually each bit of a n-bit bus.

The benefits are multiple. First, it would reduce the signal congestion  observed with wide buses, which limits the frequency.
Second, both the interconnection cost and the configuration memory footprint would be reduced because a single configuration bit would control a n-bit bus. Hard-wired wide buses would also ease the routing process for the FPGA tools.

While many applications using byte based structures would benefit from a hard-wired wide bus, single bit routing level is still required to keep the flexibility of today's FPGAs.


\section{Related Work}

\textbf{FPGA acceleration.} FPGAs can drastically improve the CPU usage efficiency, reduce the computation time, and increase the energy efficiency. \citeauthor{cloud_scale_acceleration_architecture} have demonstrated the benefits of FPGAs on applications such as machine learning or page ranking~\cite{cloud_scale_acceleration_architecture}. 
For pure network acceleration, the NetFPGA platform has been introduced. However, it lacked support for programmable data planes. Recently, \citeauthor{p4-netfpga} demonstrated how to extend the NetFPGA platform to implement programmable data planes using the Xilinx SDNet compiler~\cite{p4-netfpga}.
Moreover, many works have proposed to use FPGAs as SmartNICs~\cite{firestone,cloud_scale_acceleration_architecture}. However, these works connect FPGAs to 40G/50G Ethernet links, which do not stress the FPGA architecture.

\textbf{FPGA architecture for networking.} While \citeauthor{Bosshart:2013:FMF:2534169.2486011} highlighted that FPGAs cannot beat a programmable ASIC because of their inefficient TCAM support and the limited I/O bandwidth provided, the authors did not study the roots of these inefficiencies~\cite{Bosshart:2013:FMF:2534169.2486011}.
\citeauthor{beyond-smartnics-towards-fully-programmable-cloud} suggested augmenting the FPGA architecture with hard-wired CAMs, but their motivation lies in supporting efficient lookup operations, which differs from our proposal~\cite{beyond-smartnics-towards-fully-programmable-cloud}. 
The Xilinx's Versal ACAP~\cite{network_on_chip_versal_acap} architecture integrates a NoC used mainly to route data out of the programmable logic, but its performance to route data within the programmable logic is yet unknown. 

\textbf{Compiling network applications to FPGAs.} Available commercial and open-source P4 compilers compile P4 directly to RTL~\cite{p4fpga,p4-netfpga}.
Another avenue is to exploit a high-level synthesis tools as an intermediate step~\cite{emu}.
However, none of these works evaluated the shortcomings of the FPGA architecture for networking applications.

\section{Conclusion}
P4 is the language and PISA the architecture that made programmable data plane a reality. FPGAs, in turn, have recently played an important role for servers and network offloading in data centers. Some works have also proposed to implement PISA on FPGAs. However, little effort has been devoted to analyze whether FPGAs can efficiently implement PISA.

In this paper, we have studied how each PISA block is mapped to the existing FPGA archicture. Our analysis, supported by experiments, showed that a few PISA blocks are inefficiently implemented on FPGAs. The root of this inefficiency lies in FPGA architectures. Still, we identified a set of networking applications, and in-networking applications that are excellent matches to current FPGA devices. We also proposed to integrate network specialized hard-wired blocks, which would significantly improve performance of FPGA-based PISA switches without sacrificing the flexibility of FPGAs.


\balance

 \printbibliography[title=References,category=cited,heading=bibintoc]
\ifCLASSOPTIONcaptionsoff
  \newpage
\fi

\end{document}